\documentclass[journal=cmatex,manuscript=article,articletitle=true]{achemso}
\usepackage[version=3]{mhchem} 
\usepackage{multicol}
 
 \usepackage{graphicx}
 \usepackage{setspace}
 \usepackage{amsmath}
\usepackage{multirow}
\usepackage{dcolumn}
\usepackage{float}
\usepackage{arydshln}
\usepackage{natbib}

\usepackage{amssymb}
\usepackage{tabularx}

\author{Romain Perriot}
\affiliation{Materials Science and Technology Division, Los Alamos National Laboratory, P.O.\ Box 1663, Los Alamos, NM 87545}
\email{rperriot@lanl.gov}
\author{Blas P. Uberuaga}
\affiliation{Materials Science and Technology Division, Los Alamos National Laboratory, P.O.\ Box 1663, Los Alamos, NM 87545}
\author{Richard J. Zamora}
\affiliation{Theoretical Division, Los Alamos National Laboratory, P.O.\ Box 1663, Los Alamos, NM 87545}
\author{Danny Perez}
\affiliation{Theoretical Division, Los Alamos National Laboratory, P.O.\ Box 1663, Los Alamos, NM 87545}
\author{Arthur F. Voter}
\affiliation{Theoretical Division, Los Alamos National Laboratory, P.O.\ Box 1663, Los Alamos, NM 87545}

\title{Evidence for percolation diffusion of cations and material recovery in disordered pyrochlore from accelerated molecular dynamics simulations}

\begin{document}





\begin{abstract}
We used classical and accelerated molecular dynamics simulations to characterize vacancy-mediated diffusion of cations in Gd$_2$Ti$_2$O$_7$ pyrochlore as a function of the disorder on the microsecond timescale. We find that cation vacancy diffusion is slow in materials with low levels of disorder. However, higher levels of disorder allow for fast cation diffusion, which is then also accompanied by fast antisite annihilation and ordering of the cations. The cation diffusivity is therefore not constant, but decreases as the material reorders. The results suggest that fast cation diffusion is triggered by the existence of a percolation network of antisites. This is in marked contrast with oxygen diffusion, which showed a smooth increase of the ionic diffusivity with increasing disorder in the same compound. The increase of the cation diffusivity with disorder is also contrary to observations from other complex oxides and disordered media models, suggesting a fundamentally different relation between disorder and mass transport. These results highlight the dynamic interplay between fast cation diffusion and the recovery of disorder and have important implications for understanding radiation damage evolution, sintering and aging, as well as diffusion in disorder oxides more generally.
\end{abstract}

\section{Introduction}
Complex oxides are critical components of numerous materials applications. Whether as solid oxide fuel cells~\cite{Jacobson2010}, supercapacitors~\cite{Mai2011}, memristors~\cite{Waser2007}, or matrices for nuclear waste encapsulations~\cite{Weber1998}, the versatility of structures and chemical compositions of complex oxides has been exploited to maximize desired properties. Mass transport plays a key role  for many of these applications, through for instance ionic conductivity and radiation damage evolution. In addition, diffusion dictates sintering~\cite{Chen1997} and aging~\cite{Tu2004} of the complex oxide materials.

Pyrochlores are a class of complex oxides with the formula A$_2$B$_2$O$_7$, where A is usually a rare earth (Gd$^{3+}$, La$^{3+}$, Lu$^{3+}$, etc.) and B a transition metal (Ti$^{4+}$, Mo$^{4+}$, Zr$^{4+}$, etc.). Among the promising properties offered by pyrochlores, amorphization resistance~\cite{Sickafus2000, Sickafus2007} and fast ion conduction~\cite{Heremans1995, Wuensch2000} stand out. The former was linked to the ability of pyrochlores to disorder by forming antisites A$_B$ and B$_A$ ($i.e.$ an A cation on a B site and vice-versa), and, along with the occurrence of natural uranium-bearing analogues\cite{Lumpkin1985}, motivates the use of pyrochlores as nuclear waste encapsulation matrices~\cite{Begg1998,Ewing2004,Perriot2016}. Ionic conductivity, in contrast, arises from the structural characteristics of the material: the pyrochlore  structure is related to the parent fluorite (BO$_2$) structure, only with distinct A and B sublattices and vacant oxygen sites to account for the reduced valence. While these structural vacancies are immobile or very slow in the ordered material, cation disorder activates these carriers, leading to a tremendous increase of the ionic conduction~\cite{Wilde1998b,Williford1999b,Perriot2015}. Experimentally, it was notably found that disordered pyrochlores can show an increase in their ionic conductivity by up to four orders of magnitude in the Gd$_2$(Ti$_x$Zr$_{1-x}$)$_2$O$_7$ system, with the disorder level increasing with the fraction of Zr introduced~\cite{Moon1988, Tuller1995}. Since the disorder is the key to fast ion conduction, it is essential to be able to predict the material's cation evolution to understand ionic conductivity.

Thus, for both these applications, understanding the dynamics of cations is critical. Further, cation diffusion governs processes such as sintering and aging. However little is known about the dynamics of cations. From a simulation perspective, this is due in part to the difficulty of reaching timescales relevant to cation diffusion while accounting for the much faster oxygen diffusion. For this reason, earlier work focused the properties of defect complexes~\cite{Li2012a} or isolated cation defects~\cite{Uberuaga2015} in otherwise ordered material, where oxygens are mostly immobile. Additionally, the role of cation Frenkel pair defects~\cite{Chartier2005} and cation interstitials~\cite{Chartier2009} on the radiation tolerance of pyrochlores was investigated by Chartier $et\ al.$, while Devanathan $et\ al.$ probed the role of the chemistry~\cite{Devanathan2010,Devanathan2013}. 

In the present paper, we used molecular dynamics (MD) and parallel trajectory splicing (ParSplice)~\cite{Perez2016}, one of the accelerated molecular dynamics methods~\cite{Voter2002}, to simulate cation vacancy diffusion in disordered Gd$_2$Ti$_2$O$_7$ (GTO) pyrochlore on the microsecond timescale. We find that a cation vacancy exhibits slow diffusion until a critical level of disorder is reached, at which point fast diffusion is triggered. Along its path, the vacancy can eliminate pairs of antisites, resulting in a concurrent decrease both in the disorder and, consequently, the diffusivity. The  migration behavior of the cation vacancy also exhibits signatures of percolation diffusion, such that the relationship between disorder and migration is complex: low disorder does not affect the diffusivity, while high levels of disorder creates a percolation network for fast cation diffusion, and increases the probability of antisite annihilation, which in turn reduces the diffusivity.

\section{Methodology}
\subsection{Molecular dynamics and disorder measure}

\begin{table}
\centering
\caption{Parameterization of the Buckingham potential used in this study, from Ref.~\cite{Minervini2000}}
\begin{tabular}{ l c c c}
\hline
\hline
Interaction  & A (eV) & $\rho$ (\AA) & C (eV/\AA$^{6}$) \\
\hline
O$^{2-}$--O$^{2-}$ & 9547.96 & 0.2192 & 32.0 \\
Gd$^{3+}$--O$^{2-}$ & 1885.75 & 0.3399 & 20.34 \\
Ti$^{4+}$--O$^{2-}$ & 2131.04 & 0.3038 & 0.0 \\
\hline 
\hline
\end{tabular}
\label{table:buck}
\end{table}

The atomic interactions of the Gd-Ti-O system are modeled by the Buckingham potential:

\begin{equation}
V(r)=\rm{A}\times \rm{exp}\left(-\frac{r}{\rho}\right) - \frac{\rm{C}}{r^6},
\end{equation}

\noindent where A, C, and $\rho$ are adjustable parameters. Formal charges are adopted, and the Coulomb interaction is calculated via the Ewald sum technique. The parameterization of Minervini et al.~\cite{Minervini2000} was adopted, as shown in Table~\ref{table:buck}. Although the Buckingham potential form is rather simple, and notably does not allow for charge transfer between ions, it has been widely used to study the properties of complex oxides, and has been repeatedly demonstrated to reveal physically meaningful trends in these types of materials, in agreement with experimental~\cite{Sickafus2000, Sickafus2007} and density functional theory~\cite{Uberuaga2004, Uberuaga2012} results. 

The MD simulations are performed with the LAMMPS code~\cite{Plimpton1995}, in the NPT ensemble at zero pressure and 3500 K. This temperature is significantly higher than the melting temperature for GTO. However, due to the relatively small size of the cells used in this work, the periodic boundary conditions (PBC) prevent melting of the material, and the temperature simply speeds up the cation transitions. Each sample consists of 2$\times$2$\times$2 unit cells of pyrochlore ($\sim$ 20$\times$20$\times$20 \AA$^3$,  704 atoms), in which A and B cations are randomly swapped to create the desired level of disorder. The resulting samples, with chemical formula (Gd$_{1-x}$Ti$_{x}$)$_2$(Gd$_x$Ti$_{1-x}$)$_2$O$_7$, with $0<x<0.5$, have a percent disorder level of:

 \begin{equation}
 y=2\times x\times100.
 \end{equation}
 
\noindent In a fully disordered case (100\%), the occupancy on each cation crystallographic site is half A and half B, and the disorder is maximal (2$\times  0.5\times100=100\%$). This measure of disorder via site mixing is similar to that employed in experimental studies~\cite{Zhang2015}. In the following, the level of disorder is measured through the so-called reference lattice method~\cite{Uberuaga2014}, whereby antisites are identified through comparison with a reference perfect lattice. As shown in the supplementary information, our conclusions also hold when 
a different measure of disorder, one which would correctly quantify disorder in samples that contain ordered domains, is used instead.
Before the simulations begin, a Ti cation vacancy is introduced in the sample, which is then equilibrated at the desired temperature by running for 10000 steps (10 ps).

\subsection{ParSplice}

As will be shown below, disorder promotes fast cation diffusion so that  direct MD simulations are adequate at high disorder; cation diffusion is however considerably slower
at low disorder. In this regime, we instead rely on the parallel trajectory splicing (ParSplice~\cite{Perez2016}) method.
ParSplice is an extension of parallel replica dynamics (PRD~\cite{Voter1998, Perez2015}). In PRD, the simulation is parallelized in the time domain by simultaneously exploring a given state with multiple replicas. However, in the standard PRD method, states are visited sequentially. Therefore, the efficiency of a PRD simulation, and thus the number of processors that can be applied, is limited by the escape time out of individual states.  ParSplice overcomes this limitation by exploring multiple states at once.  If the state-to-state trajectory enters one of these states,  the pre-calculated segments of trajectory can be directly spliced onto the existing one. Segments are carried out in different states according to the estimated likelihood that they will be visited by the trajectory in the future;
this likelihood is obtained from a kinetic model that is updated on the fly as the simulation proceeds.  
Note that speculation with regard to the states that will be visited in the future only serves to allocate ``excess'' resources that could not be efficiently leveraged by a traditional PRD approach; ParSplice therefore always outperforms PRD, especially so for systems containing sets of states that are revisited many times before the trajectory moves on to a different region of configuration space (so-called  ``superbasins'')~\cite{Perez2014} .

In the ParSplice simulations, states are defined based on the position of cations alone, $i.e.$, a state-to-state transition is deemed to have occurred only when there is a change on the cation lattice. This is justified by the fact that fast oxygen motion allows for a rapid equilibration with respect to the instantaneous cation configuration.
Computational studies on La$_2$Zr$_2$O$_7$ notably showed that the oxygen sublattice quickly (0.5~ps) responds to an imposed cation disorder~\cite{Crocombette2007}, a result that should apply for other pyrochlore chemistries. This was explicitly verified here by demonstrating that the waiting times for a cation jump event
is an exponentially distributed random variable, indicating a sufficient separation of timescales between cation and oxygen dynamics. Details are given in the supplementary information.

\section{Results}
\begin{table*}
\centering
\caption{Summary of PRD and ParSplice simulations of cation vacancy migration, showing the initial and final number of antisites (and the associated disorder level), the number of transitions ($i.e.$ vacancy moves), and the total simulation time reached. Only the first listed simulation, with disorder $y=0\%$, was performed with PRD. \label{tab:parsplice}}
\begin{tabular}{ l c c c}
\hline
\hline
N$_{anti}$[initial] (disorder) & N$_{anti}$[final] (disorder) & Number of transitions & simulation time ($\mu$s)  \\
\hline
0 (0.0\%) & 0 (0.0\%)  & 360 & 3.3 \\
9 (7.0\%) & 9 (7.0\%)  & 1065 & 4.5 \\
17 (13.3\%) & 15 (11.7\%)  & 5303 & 15.2 \\
24 (18.8\%) & 23 (18.0\%)  & 3549 & 5.2 \\
39 (30.5\%) & 29 (22.7\%)  & 5904 & 3.3 \\
51 (39.8\%) & 35 (27.3\%)  & 9562 & 2.8 \\
65 (50.8\%) & 39 (30.5\%)  & 7977 & 1.9 \\
\hline 
\hline
\end{tabular}
\label{table:parsplice}
\end{table*}

Using the methods described above, we simulated the diffusion of a Ti vacancy in GTO. For low disorder levels, ParSplice simulations are needed in order to reach a significant number of cation moves. ParSplice simulations were performed for initial disorder levels between 6.25\% and 50\% (see Table~\ref{tab:parsplice}). Results show that the Ti vacancy immediately decays into V$\rm_{Gd}$+Ti$\rm_{Gd}$, as was previously observed in ordered Lu$_2$Ti$_2$O$_7$\cite{Uberuaga2015}, such that the resulting initial number of antisites is higher by one than initially constructed. For higher levels of disorder, cation diffusion is significantly enhanced, and we use traditional MD simulations instead. Since these simulations require considerably less computer resources than ParSplice runs, we perform five simulations for each level of disorder, using different structures that are all representative of the material at a given level of disorder.

Figure~\ref{fig:D} summarizes the results obtained from the ParSplice and MD simulations for the cation diffusion due to a cation vacancy in GTO. 
The diffusion coefficients were extracted from the mean-squared atomic displacements (MSD) averaged over entire simulations. It should be noted that, since the disorder evolves during the simulation (see below),
the diffusion constant is not measured in equilibrium conditions. The ``non-equilibrium diffusion coefficient'' we measure is therefore denoted by D$^*$, to differentiate it form an equilibrium diffusion constant D.  Note, however, that the change in disorder during a typical simulation is small compared to the change required to induce a large change in the diffusivity.
 For zero disorder, two additional approaches are employed to determine the diffusion constant. First, as the cation lattice is ordered and thus there is no superbasin (the oxygen ions do not diffuse in ordered GTO), we can use standard PRD to simulate cation migration. Second, the diffusion constant can be described by an Arrhenius relation:

\begin{equation}
\rm D=D_0\times exp\left(\frac{-E_a}{k_B T}\right),
\end{equation}

\noindent where D$_0$ is the temperature independent prefactor, E$_a$ the activation energy, k$_B$ the Boltzmann constant and T the temperature. E$_a$ is obtained by performing nudged elastic band (NEB\cite{Henkelman2000}) calculations for the migration of a Gd vacancy.  The prefactor D$_0$ is:

\begin{equation}
\rm D_0=\frac{1}{2N}\times f\times S^2\times Z\times  \nu_m,
\end{equation}

\noindent with $N$ the dimensionality of the system, $f$ the correlation factor (here approximated as 1), $S$ the jump distance (3.6 \AA), and $Z$ the number of neighboring sites the vacancy can hop to (3 of the same chemistry). The attempt frequency $\nu_m$ is obtained via calculation of the normal modes of vibration (Vineyard method~\cite{Vineyard1957}):

 \begin{equation}
 \nu_m=\prod_{n=1}^{3N-3}\nu^{initial}_n\bigg/\prod_{n=1}^{3N-4}\nu^{saddle}_n,
  \end{equation} 

\noindent with $\nu^{saddle}_n$ and $\nu^{initial}_n$ the normal modes at the saddle and the initial state, respectively. The three translational modes are excluded, as well as the additional imaginary mode characterizing the saddle. In this work, we obtained E$_a=4.84$ eV and $\nu_m=1.04\times10^{13}$~Hz for a Gd$\rightarrow\rm V_{Gd}$ migration.

From Fig.~\ref{fig:D}, we see that the diffusion constant is insensitive to the disorder level until $y\sim25\%$, after which it rapidly increases (by two orders of magnitude). This increase in diffusivity appears to slow down above $y\sim50\%$, but it is difficult to assess whether the diffusivity fully saturates close to full disorder.
The horizontal bars on the figure represent the change in disorder during the course of the given simulation. As discussed below, as the vacancy migrates through the material, it can enable the annihilation of antisites and hence promote the reordering of the material. Importantly, the amount of reordering depends on the given simulation, even for the same initial level of disorder. Also, for a given level of disorder, different simulations result in a different value of, and thus a spread in, the diffusion constant. These two observations highlight the effect of the local cation environment on the cation mobility: depending on the local cation arrangement, both the rate of migration and the rate of antisite annihilation can be relatively fast or slow. 

To illustrate the dynamical nature of cation diffusion in the disordered material, additional details from a particular case, the ParSplice simulation with starting disorder 50.8\%, are shown in Fig.~\ref{fig:x50}. The evolution of the total cation squared displacement shows a trend for the rate of displacement to decrease as the simulation progresses, as indicated by the slopes at the beginning and the end of the simulation. 
Further, the evolution of the number of antisites show that it does not steadily decrease. Rather, bursts of fast antitisite annihilation are followed by long periods when the number of antisites remains constant. In addition, the average rate of antisite annihilation seems to decrease as well, with 16 antisites annihilated in the first 0.25 $\rm\mu s$, but only 10 in the remaining 1.5~$\rm\mu s$. 
After 1.8~$\rm\mu s$, the total number of antisites has decreased by 20\%, bringing the disorder level down to 30.5\%. Thus,  the rate of vacancy diffusion and the rate of antisite annihilation appear to be intimately coupled. 

\begin{figure}
\includegraphics[width=0.7\columnwidth]{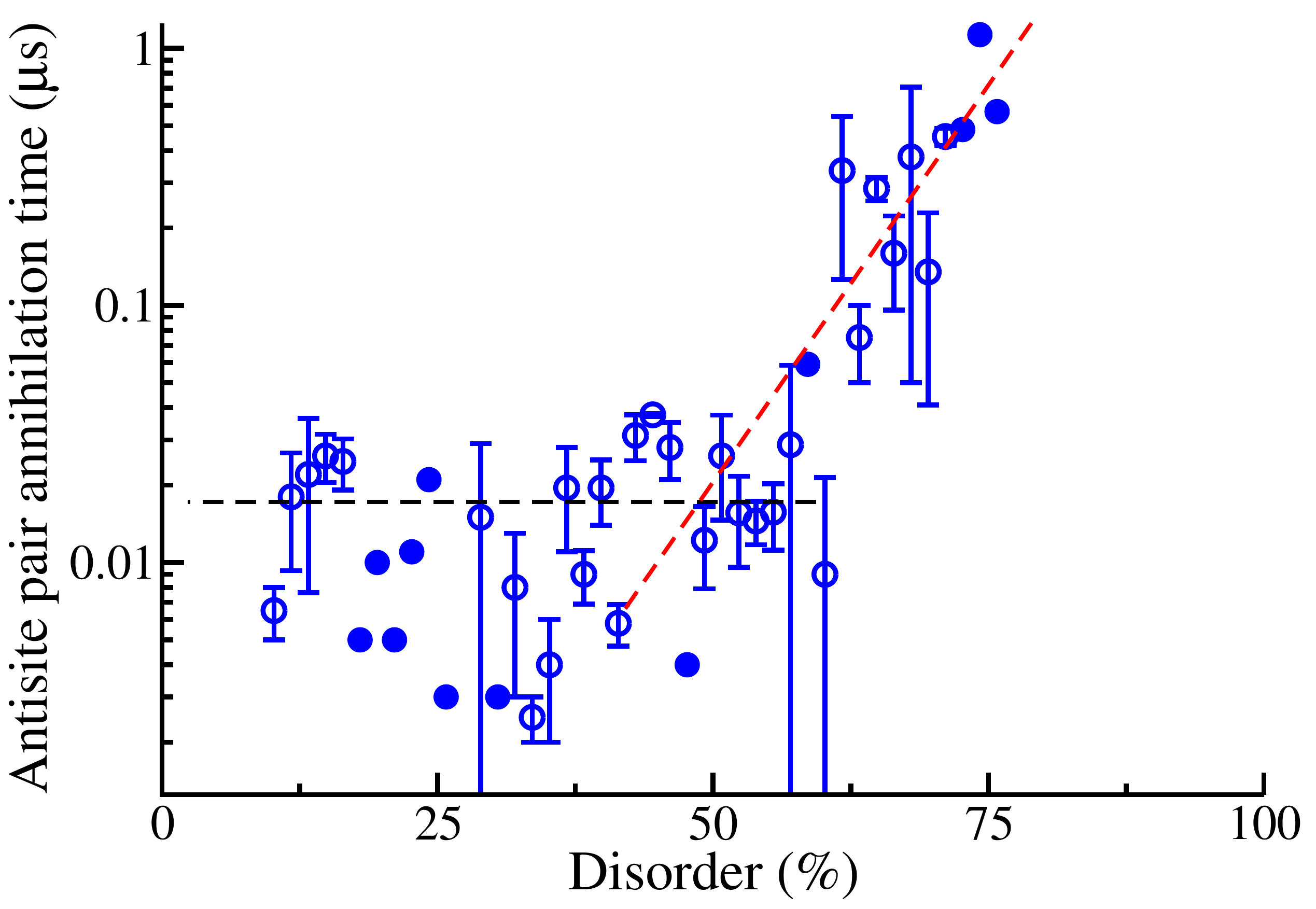}
\caption{Time to annihilate antisite pair, as a function of the number of antisites in the system. Closed symbols denote cases for which only one data point is available, $i.e.$ the transition n$\rightarrow$n-1 antisites was observed only once over all simulations. Error bars reflect the standard error for averaged values when more than one data point was available. The red dashed line is obtained by fitting an exponential function, and the black dashed line by fitting to a constant.\label{fig:disorder-change}}
\end{figure}

To further illustrate the relationship between the level of disorder and the rate of reordering (antisite annihilation), we analyzed the rate of antisite annihilation as a function of disorder as observed in all the simulations (MD and ParSplice); this data is compiled in Fig.~\ref{fig:disorder-change}. It is worth mentioning that finding the reference corresponding to a fully disordered sample is not trivial. Indeed, the lattice reference method implies that one can clearly identify the ordered and disordered domains of the material, which is not straightforward for a fully disordered system. Here, we shift the reference structure onto all cation sites of the disordered sample, and look for the minimum mismatch ($i.e.$ number of antisites). Over the course of the material evolution however, since the reordering happens at the local level, it may happen that the minimum mismatch configurations differ for the beginning and end of the run; in this case, tracking the disorder evolution is ambiguous and such points were excluded from the data in Fig.~\ref{fig:disorder-change}.

It can be seen from Fig.~\ref{fig:disorder-change} that the time necessary to annihilate an antisite pair quickly drops with increasing numbers of antisites in the system. There are two main reasons for this: first, a higher vacancy mobility increases the rate at which the vacancy can find and annihilate antisites, and second, a higher concentration of antisites leads to a higher concentration of antisite {\em pairs}. 
This last point is crucial as antisite annihilation occurs through the reaction: 

\begin{equation}
\rm{V}_{Gd} + (\rm{Gd}_{Ti}+\rm{Ti}_{Gd})\rightarrow \rm Gd_{Gd}+ {V}_{Ti}+\rm{Ti}_{Gd}\rightarrow Ti_{Ti}+\rm{Gd}_{Gd} + \rm{V}_{Gd}.\label{eq:annihilV}
\end{equation}

\noindent If opposite antisites are not neighbors, the V$_{Ti}$ in the second step of the reaction in Eq.~\ref{eq:annihilV} has to diffuse until it finds the second antisite.
However, as mentioned previously, V$_{Ti}$ is not kinetically stable in GTO and cannot diffuse through the material. Thus, if Gd$_{\rm Ti}$ is not a neighbor of Ti$_{\rm Gd}$, the second step will essentially reverse, reverting back to V$_{\rm Gd}$+Ti$_{\rm Gd}$. The two fitting lines in Fig.~\ref{fig:disorder-change} illustrate that two regimes of antisite annihilation (slow and fast)  appear to exist whether the disorder is  below or above 50\%.

\begin{figure}
\includegraphics[width=0.5\columnwidth]{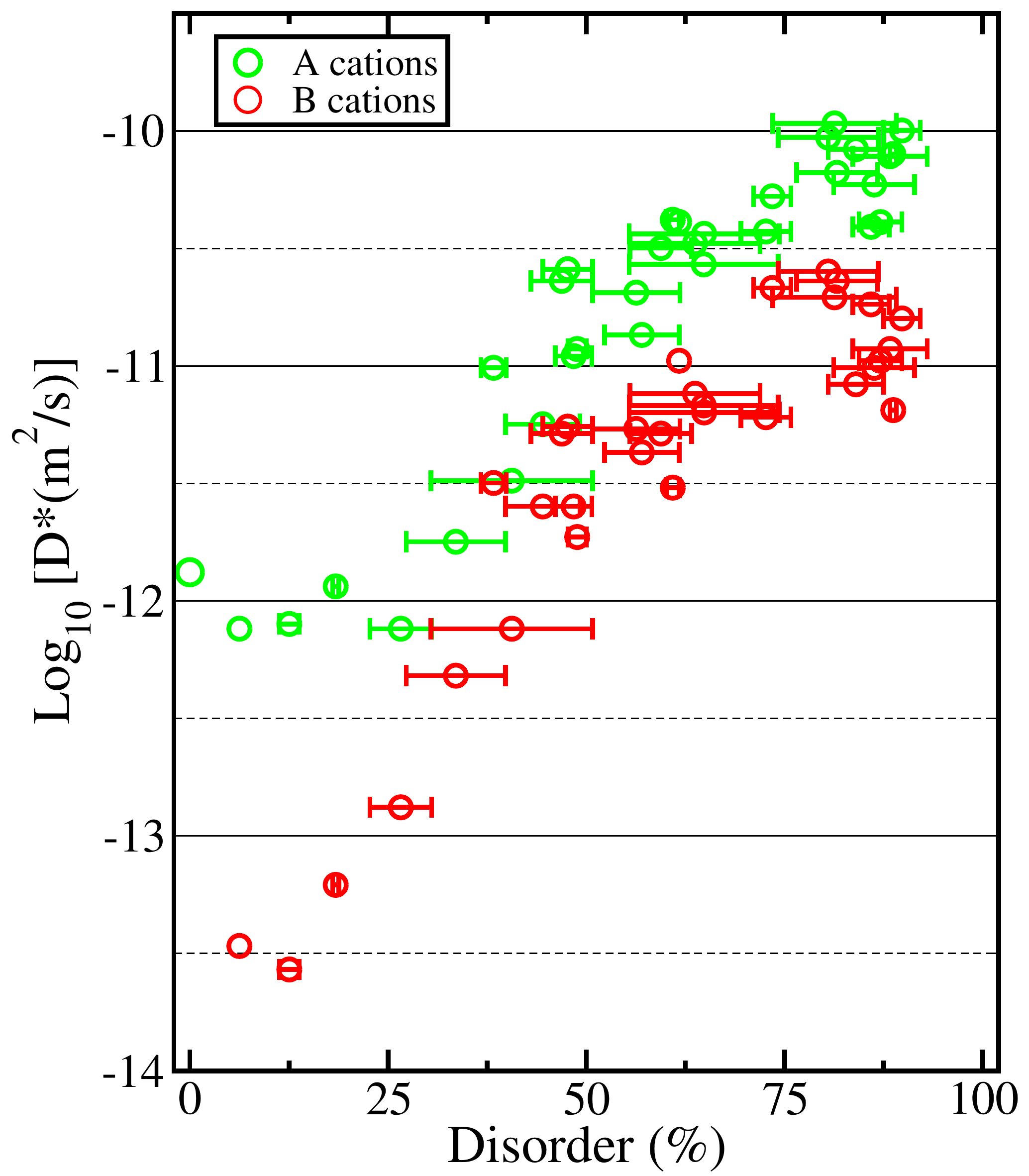}
\caption{Non-equilibrium diffusion coefficient D* (see text) of A (green) and B (red) cations due to vacancy-mediated diffusion as a function of disorder in the material. Points obtained by PRD, ParSplice, and MD.\label{fig:AB}}
\end{figure}

Note that the diffusion of the vacancy can itself promote the formation of the antisite pairs that are required for ordering. While the fact that the vacancy remains in the $\rm V_{Gd}$ form most of the time limits the diffusivity of the cations on the B lattice, our simulations show that V$\rm_{Gd}$+Ti$\rm_{Gd}\Leftrightarrow$V$\rm\rm_{Gd}$+Ti$\rm_{Gd}$ reactions in which V and Ti swap Gd sites are 
possible, if less probable than V$_{\rm Gd}$+Gd$_{\rm Gd}\Leftrightarrow$V$_{\rm Gd}$+Gd$_{\rm Gd}$ reactions. As shown in Fig. ~\ref{fig:AB}, in which the individual A and B contributions to the overall cation diffusion are separated,  
the diffusivity is dominated by A (Gd) cations (also compare with Fig.~\ref{fig:D}). The diffusivity of B (Ti) cations is initially very low, but steadily increases with disorder, as 
a result of the process described above. Note that antisite annihilation also contributes to the B diffusivity, especially at higher disorder.

\section{Discussion}

\begin{figure*}
\includegraphics[width=0.9\columnwidth]{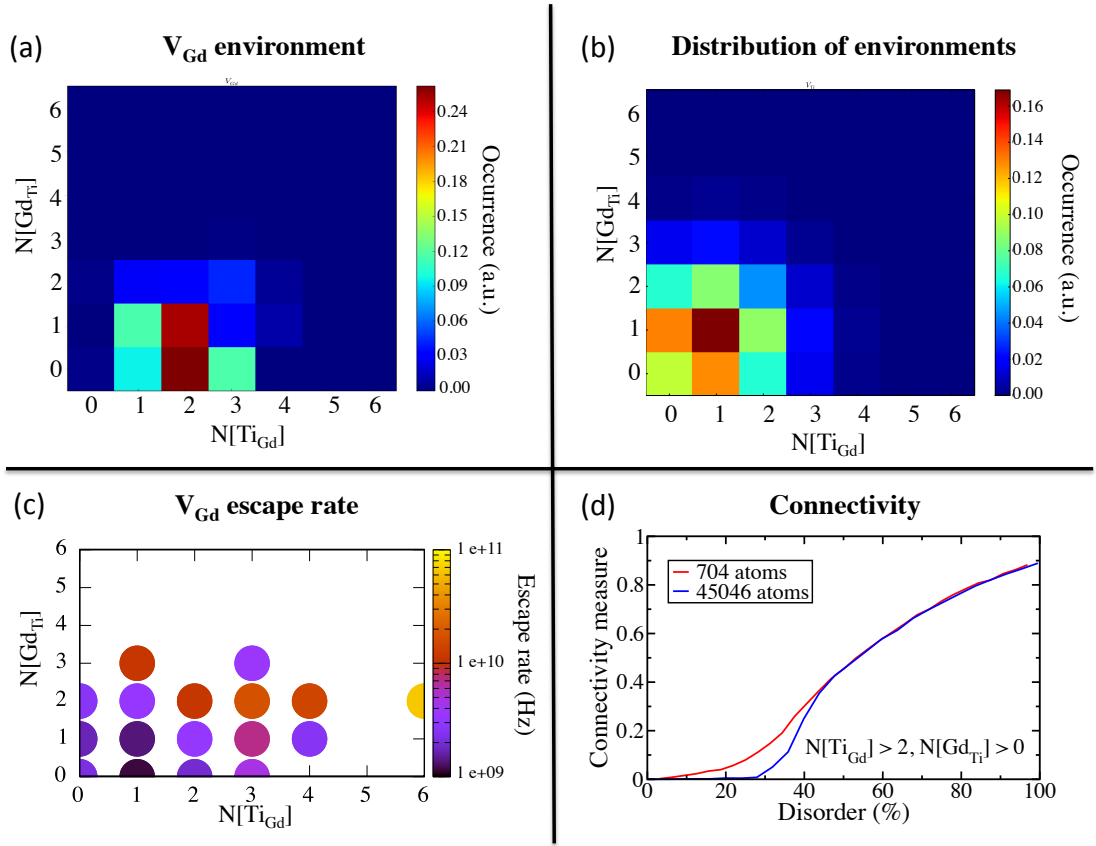}
\caption{Evidence for percolation diffusion in disordered pyrochlores. (a) Preferred cation environment for the diffusing vacancy during the ParSplice simulation with initial disorder 50.8\%. The color reflects the total time spent by the vacancy in a given environment, normalized such that t$_{total}$=1. (b) Average distribution of antisite environments in a material with disorder $y$=35\%, corresponding to a typical configuration in the simulation in (a). (c) Average escape rate for the vacancy as a function of local environment for the simulation in (a). (d) 
Relative size of the largest connected components of Gd sites with at least 2 Ti antisites neighbors. (corresponding to the site most visited in (a)). The size is normalized by the total number of Gd sites in the material.\label{fig:panel}}
\end{figure*}

A key result from our MD and AMD simulations is that there is a threshold level of cation disorder, $y=25\%$, at which there is a sudden increase in the overall cation mobility. This threshold and the shape of the curve in Fig.~\ref{fig:D} 
suggests the occurrence of a percolation network of antisites that facilitate cation vacancy migration above this level of disorder, a hypothesis that is validated in Fig.~\ref{fig:panel}. Several elements are necessary for such an explanation to hold.
Firstly, the diffusing species -- the cation vacancy -- must bind, or have an energetic preference for, the structural elements that define the percolation network. Figure~\ref{fig:panel}a shows the first-nearest-neighbor environment of the Gd vacancy as it migrates during the ParSplice simulation in the sample with initial disorder $y$= 50.8\%. The results clearly reveal a preference of the vacancy for environments with 2 Ti antisites and 1 Gd antisite (2,1), and 2 Ti antisites and 0 Gd antisites (2,0). Importantly, this preference is not simply a reflection of the frequency of those environments within the structure. 

\begin{figure}
\includegraphics[width=0.75\columnwidth]{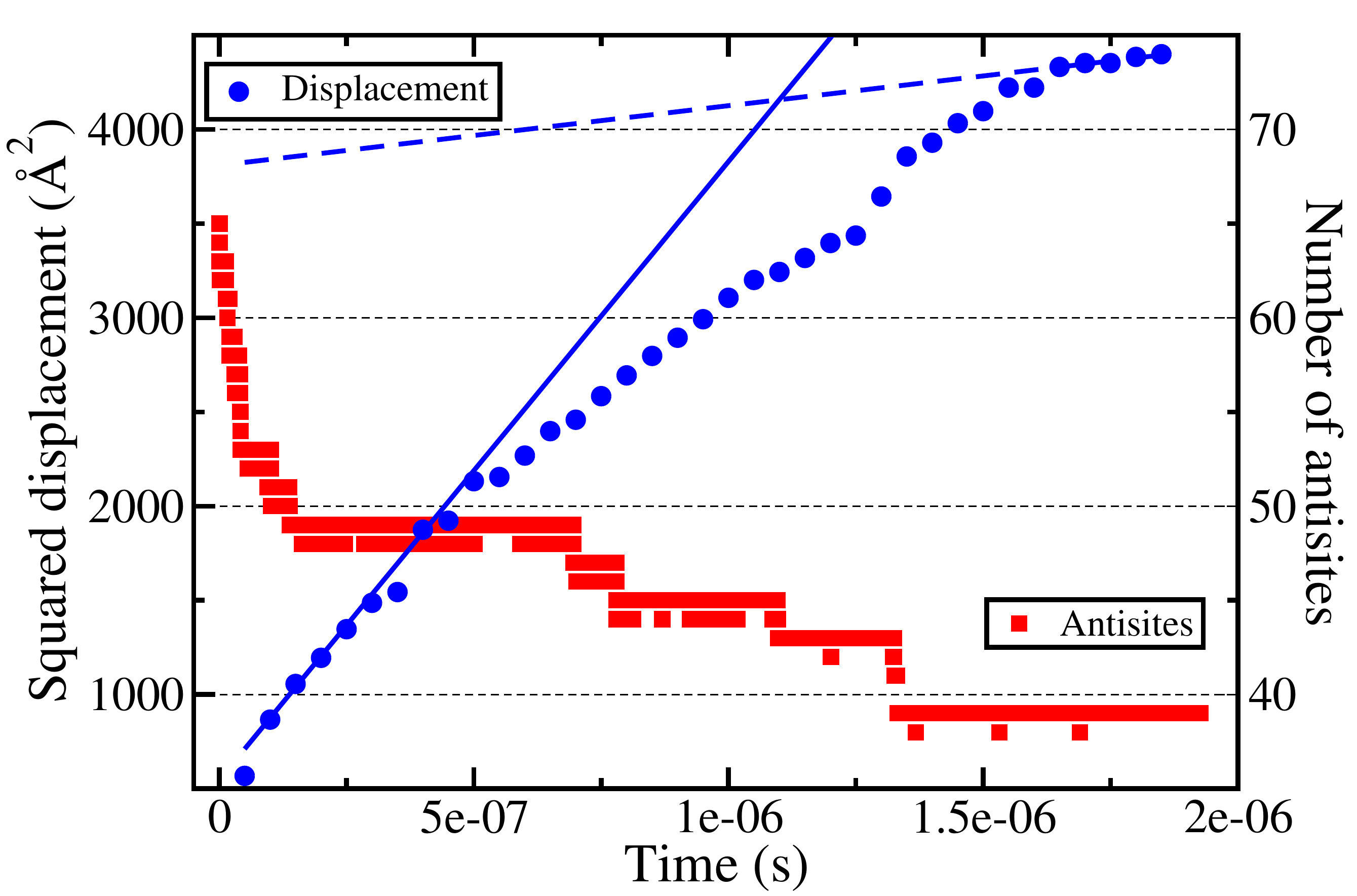}
\caption{Evolution of the number of antisites and squared displacement over time during a ParSplice simulation of cation vacancy diffusion in a sample with 65 antisites initially. The solid and dashed lines indicate the rate of change of the displacement at the beginning and end of the simulation, respectively.\label{fig:x50}}
\end{figure}

Importantly, Figure~\ref{fig:panel}b shows the distribution of cation environments in randomly generated samples at a disorder level of 35\% (this particular value of disorder was chosen as representative of the state of the system during most of the simulation, as shown in Fig.~\ref{fig:x50}). Clearly, the preference of V$_{\rm Gd}$ for the (2,1) and (2,0) environments is much higher than their representation in the structure; these sites represent only 15\% of all the sites in the material, but the vacancy spent over 50\% of the time near them. However, a preference for certain sites is not enough to lead to an enhancement of the diffusivity: there also has to be relatively fast migration between those sites. 

Further, Figure~\ref{fig:panel}c shows, for the same simulation, the dependence of the escape rate of the vacancy  on its local environment. Here, it can be seen that the escape rate of the vacancy from a given site increases with the number of antisites around it. In other words, the vacancy quickly diffuses out of sites surrounded by disorder. Combined with the data in Fig.~\ref{fig:panel}a, this means that the vacancy spends more time in sites surrounded by antisites and that the rate at which it moves between these sites is high. Further, the rate to move to other sites surrounded by few antisites has to be low; 
otherwise the distribution shown in Fig.~\ref{fig:panel}a would not strongly favor antisite-rich environments. 

Lastly, Fig.~\ref{fig:panel}d displays the relative size of an antisite network as a function of the level of disorder in the material. The curves were generated by analyzing random distributions of antisites for a given disorder level and computing the fraction of A sites that are contained within the largest connected region of first-neighbor A sites with at least 2 $\rm Ti_{Gd}$ neighbors. Two sizes of the system were considered: 704 atoms (the system used in all our simulations), and 45046 atoms (2$\times$2$\times$2 times the original system), to approach the infinite limit.
As can be seen, the relative size of the largest connected clusters quickly increases above $y$=25\%, which is a finite-size signature of percolation (in an infinite system, this measure would jump from 0 to a finite value at the percolation threshold). This is the same level of disorder at which we observe cation mobility to start to increase in our simulations (Fig.~\ref{fig:D}). From the larger system, we determine the percolation threshold to occur around 35--37\% disorder.

To summarize our analysis of the percolation behavior in this system:
\begin{itemize}
\item the diffusing species ($V_{Gd}$) strongly favors sites surrounded by antisite disorder
\item the hopping rate between such sites is high
\item the connectivity of these sites increases quickly above $y$=25\% disorder
\end{itemize}
Together, these observations lead to the conclusion that vacancy-mediated cation diffusion is indeed enhanced by the occurrence of a percolation network of antisites for disorder levels above $y$=25\%. 

Remarkably, the existence of a percolation transition in the diffusion of cations is in marked contrast with what is observed for oxygen, where the diffusion steadily increases with disorder~\cite{Perriot2015}. This is likely a consequence of the high number of carriers that dictate oxygen diffusion in disordered pyrochlore, necessitating larger domains of order and disorder to form a percolation network. Indeed, we observed trapping when intrinsic (non-structural) oxygen defects were simulated~\cite{Perriot2015}. Our results are also contrary to observations in spinel~\cite{Uberuaga2015a} and perovskites~\cite{Parfitt2011,Taskin2016}, where the oxygen vacancy diffusivity is hindered by cation disorder. While in perovskites, cation ordering was shown to open particular channels for fast diffusion~\cite{Uberuaga2015b, Taskin2016}, the effect of ordering in spinel is less clear. One reason for the different behavior could be that the antisite network was fundamentally different in the simulations from Ref.~\cite{Uberuaga2015a}, either because of the way it was generated (using Monte Carlo in that study versus random generation here) or the crystal structure and stoichiometry of the two materials leads to different relationships between antisite environments. Further work is needed to understand these differences. However, it is very clear that cation disorder in complex oxides can lead to very different effects on diffusion, depending on a number of factors.

Finally, there is a vast literature on diffusion in disordered media, with various models proposed to describe such systems, from random barrier and random trap models, to models that combine both features (see Ref.~\cite{Kehr1996} for a brief summary and references therein for more detail). Our results suggest yet another type of model for cation diffusion in disordered pyrochlore. Antisites lead to deeper (trap) states for the vacancy, but critically, also to lower barriers (faster rates) between those sites. Thus, once a percolation network is established, diffusion is enhanced because, as described above, the rate of hopping between antisite environments is faster than between other types of environments. If this faster rate were not present, the formation of a percolation network of traps would not lead to faster diffusion. This effect is similar to that observed for Li diffusion in lithium-transition metal oxides~\cite{Lee2014, Lee2015}. Another important feature of the landscape of cation vacancy diffusion in disordered pyrochlore is the time-dependent nature of that landscape. As time progresses and the cation vacancy reorders the material, traps disappear and the system becomes more homogenous. Further, that antisite annihilation in this system is exothermic naturally leads to lower barriers via the Bell-Evans-Polanyi principle. As discussed above, antisite annihilation is not necessary for high cation mobility in disordered pyrochlore, but it likely contributes to it via the lower barriers associated with annihilation.

A second key result from our study is captured by Fig.~\ref{fig:x50}:  cation diffusivity is higher in disordered pyrochlore but that 
enhanced vacancy diffusivity also drives the system to reorder. This ordering in turn leads to a slowdown of the vacancy diffusion and hence of the recovery rate. Note however that 
vacancy diffusivity is only one of the contributors to the recovery rate, the others being the concentration of pairs of neighboring antisites and a possible dependence of the rate of the $ \rm{V}_{Ti}+\rm{Ti}_{Gd}\rightarrow  \rm{V}_{Gd}$ reaction on the local environment. These additional factors could explain why the cation diffusivity and the recovery rate do not 
abruptly increase at the same level of disorder.

\section{Conclusions}
To conclude, we used molecular dynamics and ParSplice simulations to investigate cation vacancy diffusion in disordered Gd$_2$Ti$_2$O$_7$ pyrochlore. We find that the diffusion in materials with low disorder is slow, independent of disorder, and almost exclusively comprised of Gd cation diffusion. However, above a threshold of $y$=25\% disorder, the diffusivity increases sharply, along with the contribution of Ti cations. We find evidence that the fast diffusion is due to the formation of a percolation network of antisites, in which the vacancy can diffuse quickly. Lastly, we find that the vacancy can heal the disorder in the material, at a rate that increases with the amount of disorder in the material. The diffusion of cations in GTO is thus a complex phenomena, where cation disorder allows for fast diffusion in a percolation network above a certain level and thus enhances healing of the material, which in turn reduces the diffusivity and the healing rate. Together, these results highlight the complex interplay between cation diffusion and disorder in pyrochlores, and provide insights into sintering, radiation damage, and aging processes.

\section*{Acknowledgements}
This work was supported by the U.S. Department of Energy, Office of Science, Basic Energy Sciences, Materials Sciences and Engineering Division. Development and implementation of the ParSplice code were supported by LANL/LDRD project 20150557ER. This research used resources provided by the LANL Institutional Computing Program. Los Alamos National Laboratory, an affirmative action equal opportunity employer, is operated by Los Alamos National Security, LLC, for the National Nuclear Security Administration of the U.S. DOE under contract DE-AC52-06NA25396.

\section{Supplementary Information}

\subsection{State definition in ParSplice}

\begin{figure}
\includegraphics[width=0.75\columnwidth]{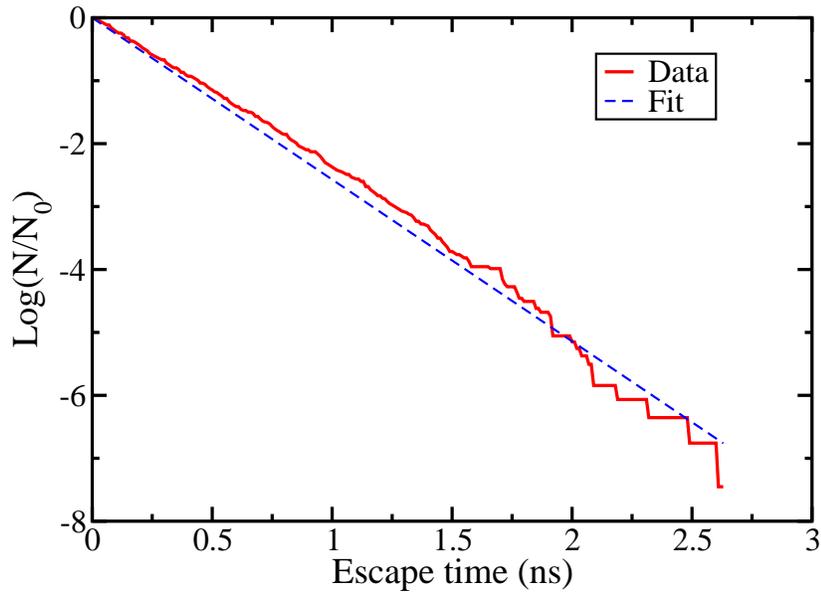}
\caption{Time for a cation vacancy in GTO to hop, or escape from its current state, at T=3000 K. $N_0$ is the total number of simulations (1724) and N is the number of simulations in which a cation jump has not been observed. The system exhibits exponential behavior (straight line fit in the log scale plot), characteristic of first order-transition behavior. \label{fig:escape}}
\end{figure}

One of the challenges in simulating the long-time behavior of this material is the fact that anions and cations diffuse at very different rates. In order to address this issue, we defined 
states in ParSplice in terms of the cations alone; $i.e.$, transitions are only deemed to have occurred if the cation lattice reorganizes; specifically, a transition is detected when the occupation of the voronoi volumes defined on the perfect lattice changes.
This definition of the states according to the cations yields accurate dynamics and efficient simulations as long as anions can be considered to rapidly achieve local equilibrium with respect to the 
instantaneous configuration of the cations \cite{Perez2014}. In this limit, the state-to-state dynamics should be described by first-order kinetics, and hence the distribution of escape 
times should be exponential \cite{Perez2014}. We verified that this condition is obeyed by running numerous independent simulations (1724 in the present case) and recording for each simulation the time at which a cation jumped to the nearby vacancy. The same initial structure with disorder level 12.5\% was used in all the simulations; however, the random seeds for the Langevin thermostat and the initial velocity assignment were randomized, creating independent trajectories. As shown in Fig.~\ref{fig:escape}, the results indeed indicate that state definition in terms of cations alone 
is warranted.

\subsection{Effect of the disorder measure}

As discussed above, the level of disorder is measured through the so-called reference lattice method~\cite{Uberuaga2014}, whereby antisites are identified through comparison with 
a reference perfect lattice. This provides a measure of site mixing relative to a reference lattice. However, one could imagine the nucleation of spatially distinct regions of ordered pyrochlore, in which case the resulting material would be composed of ordered pyrochlore domains shifted relative to one another and separated by domain boundaries. Even though such a case is unlikely in the small unit cells used here, this reference lattice method would lead to an incorrect estimation of the material's disorder, as the comparison to a reference structure could only recognize one domain as pyrochlore, while the other domains would appear as fully disordered. For this reason, a second disorder measure is employed, which identifies the ``antisite character'' of the material, by looking at the second nearest-neighbor environment of each atom. In perfect pyrochlore, the six second nearest-neighbors of a given cation are cations of the opposite species, while in a fully disordered material ($i.e.$ defect fluorite) the environment would be equally mixed (50\% A/B occupancy on each cation site). Thus, the antisite character is defined as:

\begin{equation}
\chi=\frac{n_{average}}{n_{fluorite}}\times 100\label{eq:anti},
\end{equation}

\noindent where $n_{average}$ is the average number of cations of the same species in the second nearest-neighbor shell, and $n_{fluorite}=3$. A similar measure was used in Ref.~\cite{Chartier2005} to define a cation order parameter. The antisite character (Eq.~\ref{eq:anti}), being a local measure does not depend on any reference, ensures that shifted pyrochlore domains would indeed appear to form a mostly ordered material, apart from the boundary regions. 

\begin{figure}
\includegraphics[width=0.5\columnwidth]{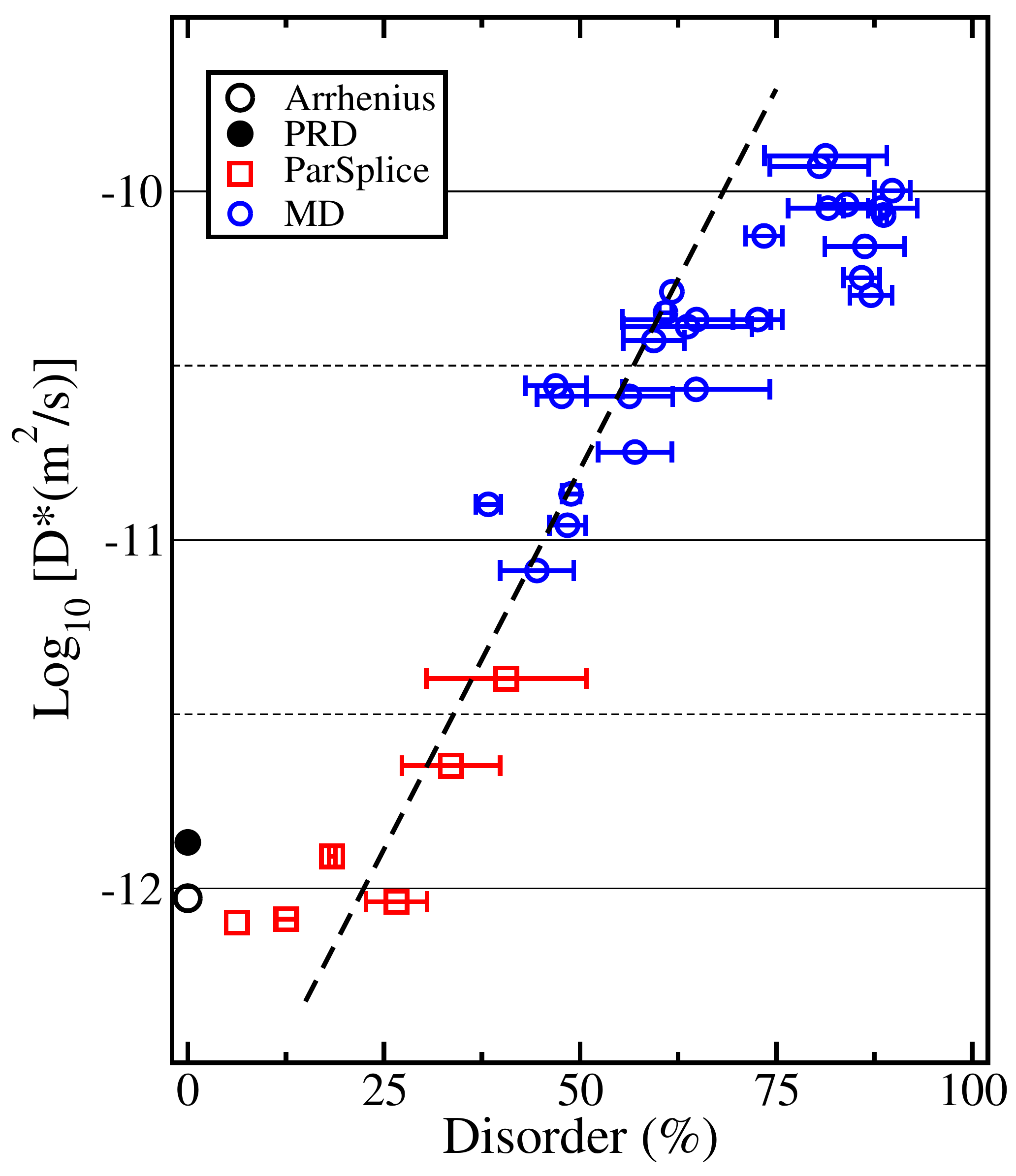}
\caption{Non-equilibrium diffusion coefficient D* (see text) for cations in GTO with a cation vacancy as a function of cation disorder. Points were obtained using PRD, ParSplice, and MD, as well as predicted from static calculations (see text).The width of the horizontal segments reflect the decrease in disorder observed in the material during the course of the given simulation. The dashed line is a linear fit to the data in the interval $y$=25--62.5\%\label{fig:D}}
\end{figure}

\begin{figure*}
\includegraphics[width=0.9\columnwidth]{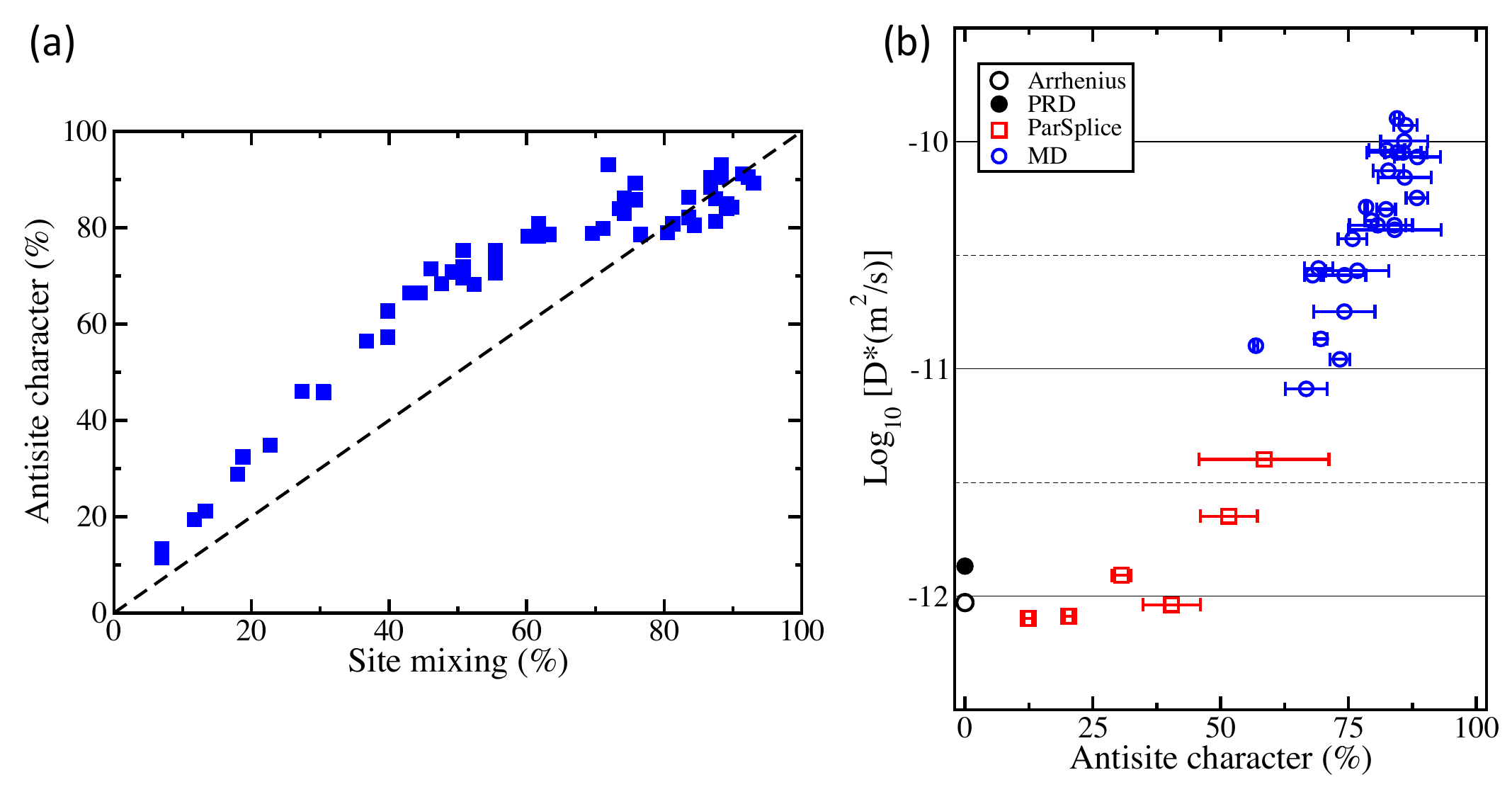}
\caption{Comparison between the antisite character and site mixing measures of disorder in the material (a), and Non-equilibrium diffusion coefficient D* (see text) for cations as a function of the antisite character in the material with a cation vacancy (b). \label{fig:D2}}
\end{figure*}

The comparison between the two measures is presented in Fig.~\ref{fig:D2}-a. It can be seen that the two measures agree at the high end of the disorder, but that the antisite character is consistently larger than the site mixing at medium and low site mixing. This is because the nature of the second nearest-neighbor environment differs for isolated antisites -- typical of low to moderate disorder and having only cations of the same species as second nearest-neighbors --  and defect fluorite -- which has a fully mixed occupancy. In addition, the antisite character is less sensitive to changes in site mixing at higher levels of site mixing.
In Fig.~\ref{fig:D2}-b, the diffusivity $vs.$ disorder results from Fig.~\ref{fig:D} are replotted with respect to the antisite character instead of the site mixing, in order to further validate the conclusions made earlier. It can be seen that the curve has a similar shape, only shifted to the right: below $\chi\sim50\%$ antisite character, slow and constant cation diffusion is observed. Above, a sharp enhancement is observed up to the maximum levels of antisite character. Further, the measured changes in disorder during the course of the various simulations is similar between the two methods, though the measure of antisite character changes less at high disorder than does the site mixing measure, a consequence of the relation exposed in Fig.~\ref{fig:D2}-a. The shift of the threshold for high diffusivity is also rooted in this relation site mixing -- antisite character.




\clearpage

\clearpage

\clearpage

\clearpage

\clearpage
\clearpage

\clearpage

\clearpage
\clearpage

\bibliography{Pu}

\end{document}